# Аномальное атмосферное явление «водяной смерч» как следствие скелетных структур океана и особый тип атмосферной аэрозольной пылевой плазмы

В.А.Ранцев-Картинов

ИЯС РНЦ "Курчатовский Институт", Москва 123182, Россия
rank@nfi.kiae.ru

Среди обнаруженных и описанных автором отдельных блоков скелетных структур океана (ССО) встречаются блоки в виде вертикально плавающих цилиндров (ВПЦ), которые грубо можно представить в виде цилиндров плотно упакованных тонкими капиллярами, ориентированными вдоль оси цилиндра. Предполагается, что эти капилляры собраны из углеродных нанотрубок третьего поколения. Выдвигается гипотеза о формировании начальной стадии аномального атмосферного явления - "водяной смерч" (ВС), который провоцируется ВПЦ ССО. Интерпретация ВС дана на основе капиллярных явлений в присутствии высокой напряженности электрического поля, обусловленного заряженным до высокого потенциала облаком над ВПЦ. Рассмотрена капиллярно – электростатическая модель начальной стадии формирования ВС, следствия которой идентичны наблюдаемым особенностям ВС. Высказано предположение о сценарии развития ВС и возможного перехода его в классический торнадо.

**1. Введение.** Анализ фотоизображений поверхности океана, полученных с различных высот и при различных уровнях волнения его поверхности, привел автора к наблюдению скелетных структур океана (ССО) [1,2]. Каркасные структуры (КС), подобные ССО с подобной топологией и с теми же свойствами (долгоживучестью, самоподобием и стремлением к самосборке) ранее уже наблюдались в целом ряде явлений в лаборатории, атмосфере и космосе [3-5]. Итоговые работы по КС, [3е,4а], дали выводы *о роли нанопыли в формировании КС в широком диапазоне масштабов*, охватывающих почти 30 порядков величины. В [1,2] автором были выдвинуты гипотезы по формированию отдельных блоков и сети ССО. Основная гипотеза гласит: *«Создавая (за счет адсорбции растворенного в воде газа) на своей поверхности разделение трех фаз вещества (твердой, жидкой и газообразной), КС обеспечивают действие сил поверхностного натяжения (в условиях неполного смачивания) и сцепления между собой своих фрагментов и отдельных блоков в единую сеть ССО даже под водой»*. Наблюдение, описание [1] и интерпретация [2] ВПЦ могут стать подтверждением данной концепции. Другая гипотеза говорит - *о роли атмосферного электричества в формировании КС таких облаков, которые могут привести к аномальным атмосферным явлениям (ААЯ)*, и напрямую связана с темой данной работы.

**2. Наблюдения "водяного смерча".** Согласно Д.Фрэзеру [5], традиционной дефиницией "водяного смерча" (ВС) является ВС – это торнадо, существующий над большой водой. ВС имеет много отличий от классиче-ского торнадо, например, он: a) существует всего 15-30 минут; b) в 5 раз меньше в диаметре, скорость движения и вращения его ниже раза в 2-3; c) ветер может быть не ураганным, и волнений поверхности моря может почти не быть; d) начинается ВС от поверхности моря, а не с воронкообразного пальца в материнском облаке (МО); e) ВС может быть как циклоническим, так антициклоническим и достигать высоты 0,1-1 км; f) начальная стадия водяного смерча характеризуется появлением светящегося пятна на поверхности воды диаметром 10-20 м и темного круга диаметром в 3 раза большим вокруг него; g) в это время нет вращения и спускающегося пальца из МО. На второй стадии развития ВС: 1) появляется спираль и вращение, которое постепенно усиливается; 2) в центре светлого пятна появляется темная область и шнур становится трубчатым; 3) вращение возрастает; 4) вокруг появляется диффузное кольцо мелких капель и пара; 5) с этого момента начинается поступательное движение колонны; 6) шнур резко сжимается по радиусу и из МО вытягивается палец, как в торнадо. Это кульминация ВС. После чего наступает стадия его распада, с ослаблением скорости вращения, наклоном и вытягиванием по горизонтали основной его колонны и развалом ее на части. На рис.1 приведена иллюстрация ВС из коллекции фото архива NOAA [6].

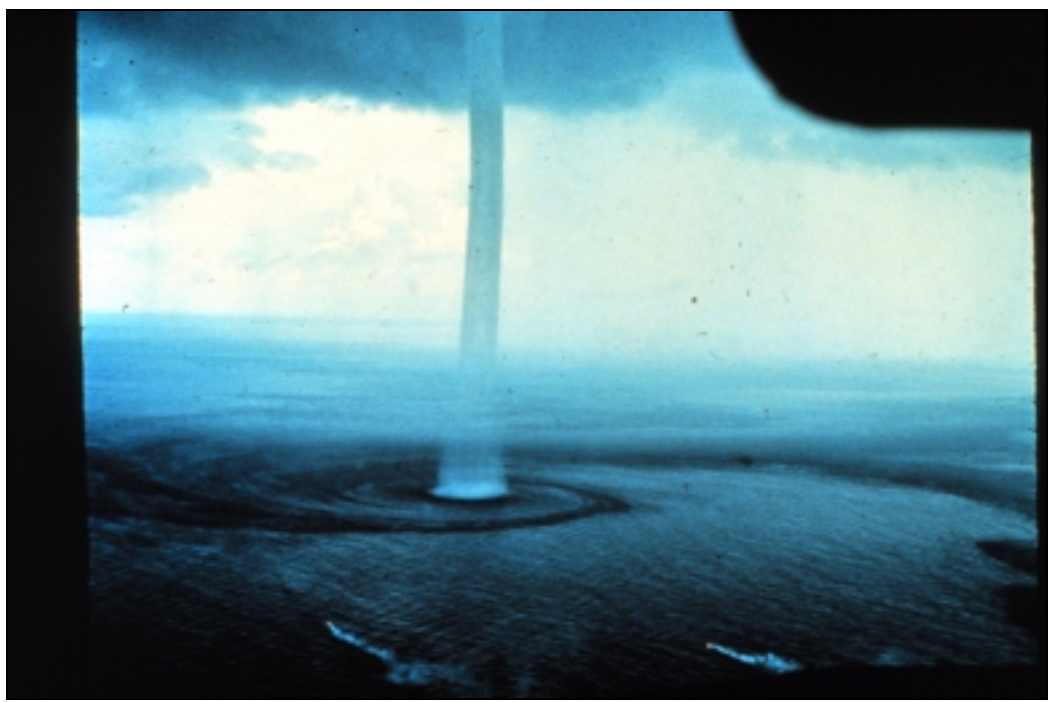

Рис. 1. Фотография водяного смерча получена летающей лабораторией во Флориде. На снимке видны почти все отмеченные выше основные признаки начальной стадии развития ААЯ водяной смерч.

**3. Интерпретация наблюдаемого явления**, При описании ССО [1] и интерпретации явления [2], особое внимание было уделено блокам в виде ВПЦ. Была предложена модель их сборки при различных уровнях колебания поверхности океана. Исходя из этой модели, были сделаны оценки их плавучести и прочностных свойств [2]. Рассмотрим случай, когда над штилевым морем появилось МО с достаточно высоким значением потенциала по отношению к поверхности океана. При высокой напряжённости электриче-ского поля у поверхности океана ССО "оживают". Особенно характерным будет поведение ВПЦ [2], которые плотно упакованы капиллярами (из углеродных нанотрубок (УНТ) или им подобных). Если один из них имеет такую плавучесть, что его торец слегка выступает из воды, то этого оказывается достаточным, чтобы спровоцировать именно его взаимодействие с заряженным МО. Помещение капилляров в сильное электрическое поле ведет к образованию цепочек аэрозольных капель, вытягиваемой из капилляров жидкости и ориентированных вдоль поля. Эти капли (благодаря эмиссии электронов с торцов капиллярных трубок) быстро заряжаются (особенно, если капилляры – УНТ, обладающие высокой термоэмиссией электронов даже при комнатной температуре). Высокая напряженность поля способствует формированию и зарядке капель на торцах капилляров, и подъему жидкости вместе с капиллярами на некоторую высоту (благодаря силам поверхностного натяжения (СПН)). Высота подъема жидкости в капиллярах определяется этими силами. На предельной высоте СПН уравновешиваются весом столба жидкости поднятого ими. Тогда для отрыва сформировавшейся капли достаточно всего небольшого усилия, поскольку у торца капилляра СПН израсходовали свою энергию. Электрическое поле должно теперь затратить работу только на подъем оторванной капли. Заряженные капли, отрываясь, устремляются вверх, унося накопленный на их поверхности заряд и замыкая электрическую цепь разрядного промежутка переносом заряда аэрозольных частиц. Но по мере нарастания мощи процесса, с водой из ВПЦ начинают вытягиваться короткие куски капилляров и даже целые его фрагменты, частично наполненные/смоченные водой. Поскольку ССО связывает свои отдельные блоки в единую сеть, то этот процесс вовлекает часть ее в свое поле активной деятельности. Это приводит к тому, что со временем из воды в жерло ВС понемногу начнут втягиваться примыкающие к этому ВПЦ фрагменты сети ССО. Все это можно отнести к разряду спокойного начального периода становления ВС. Согласно [6], ВС характеризуется формированием вертикальной и тонкой колонны. Видно, что в колонне на начальном этапе формирования ВС нет необходимости в высокоскоростном вращении массы колонны и газа. На этом этапе природа явления может быть чисто капиллярно - электростатическая. В случае же, когда в самом МО уже была хорошо развита КС, процесс формирования ВС в направлении от поверхности океана совмещается с вытягиванием части КС из МО, т.е., в направлении вниз, что соответствует начальной стадии формирования классического торнадо [3e,4a,c]. Если такой ВС просуществует долго, то это может привести к перерождению его в мощный торнадо. Действительно, в таком случае через колонну ВС в МО вносится дополнительная масса уже готовых структурообразующих фрагментов ССО с большим количеством воды. В условиях МО они способны быстро создать в нём развитую КС, способствующую более высокой концентрации энергии и массы в колонне, благодаря возможности нелокального переноса (по КС) энергии поля МО. Если такой переродившийся ВС выйдет на сушу, то разрушения его будут колоссальными, что в действительности и наблюдается в природе.

Теперь попробуем оценить, каким физическим условиям нужно удовлетворить, чтобы предложенный механизм формирования ВС мог осуществиться. Согласно [2], базовым элементом построений цилиндрических структур ССО является УНТ 3-го поколения. Её диаметр (см. [2]) $D \sim 1{,}7 \cdot 10^{-3}$ см, длина $L \sim 6 \cdot 10^{-3}$ см и собственная масса $M \sim 1{,}7 \cdot 10^{-11}$ г. По-гонная масса воды, заполняющей такой капилляр, в 1000 раз больше погонной массы самого капилляра. Поэтому в оценках мы этой массой будем пренебрегать. Если напряженности электрического поля в промежутке заряженное МО/водная поверхность достаточно, чтобы поднять капиллярную нить на некоторую высоту и удерживать её в вертикальном положении, то вода в ней поднимется за счёт СПН на высоту $h$, определяемую капиллярными силами. Будем считать, что диаметр аэрозольных заряженных частиц воды, оторвавшихся от торца капилляров $\sim D$. Оценим теперь минимальный заряд $q_{min}$ такой капли, чтобы она могла стартовать вверх, $q_{min} = \pi D^3 \rho g / 6E$. Здесь $\rho$ - плотность воды, $g$ - ускорение свободного падения, $E$ - напряженность электрического поля в промежутке МО/океан. Для нашего случая $q_{min} \sim 1,5 \cdot 10^2$ электронов. Ограничением для заряда является условие, когда произойдёт кулоновский взрыв поверхности шарика, т.е., когда потенциал заряженной поверхности этой капли превысит энергию её поверхностного натяжения. Если $q_{max}$ - максимально допустимое значение заряда поверхности капли, то $q_{max} \sim \left(\pi D^3 \sigma / 2\right)^{1/2}$, что в нашем случае дает $q_{max} \sim 10^6$ электронов. Капля воды, отрываясь от своего столбика жидкости и зацепившись СПН за край капилляра, может вынести часть его под действием электрического поля на высоту МО. Учитывая все это, можно сказать, что структура ВПЦ является хорошим спусковым крючком для вышеописанного процесса. По мере развития ВС в его жерло втягиваются фрагменты ССО вместе с аэрозольной водяной пылью. Капилляры ВПЦ не синхронно откликаются на воздействие электрического поля. На одних процесс идет быстрее, другие же будут отставать и получится некоторое стохастическое распределение подъёма, как самих капилляров, так и образующейся аэрозоли. У основания ВС над водной поверхностью будет капиллярно-водяной пьедестал с рваной верхней границей. Выше же он будет продолжаться потоком аэрозольных частиц и фрагментов сети из капиллярных блоков различных размеров и ориентации, а также пучков длинных капиллярных нитей, связанных между собой СПН водяной пленки, покрывающей их боковые поверхности. На самой границе перехода от пьедестала к аэрозольному восходящему вверх столбу должна иметь место зона повышенного рассеивания света, обусловленная образованием множества капель и воздушных пузырьков в водяной оболочке. В этой зоне должно наблюдаться некоторое дополнительное свечение из-за возбуждения молекул воды и воздуха при обрыве электрических цепей капилляров в момент образования капель аэрозоли. Оценим теперь средний ток в колонне. Положим, что расстояние между аэрозольными каплями составляет величину порядка десяти диаметров частиц (именно такие расстояния фиксируются в экспериментах по образованию капиллярами аэрозолей в электрическом поле такой напряженности). Тогда каждый капилляр может выдавать $\sim 10^5$ частиц в секунду. При плотном заполнении круга диаметром $\sim 10$ м, такими капиллярами получается $\sim 10^9$ капилляров. Следовательно, производительность сечения ВПЦ будет $\sim 10^{14}$ частиц/с. Положим, что каждая капля способна нести заряд $q \sim 10^4$ электронов (такой заряд является характерной величиной для экспериментов с пылевой плазмой). Это соответствует суммарному току $\sim 0,5$ А. Отсюда мощность этой электростатической машины W ~ 5 10⁸ Ватт, поскольку потенциал МО порядка V ~ 10⁹ В. За 20 минут работы ВС совершит работу ~ 6 10¹¹ Дж. По мере подъёма капель их заряд возрастает и, наконец, наступает момент, когда происходит кулоновский взрыв капель и их масса переходит в водяной пар. Этот заряженный пар вместе с частично ионизованным газом атмосферы со скоростью, вплоть до $V \sim 10^4$ см/с (измеряемая в ВС величина) поднимается вверх. Это ведёт к понижению давления в аэрозольной струе. Она начинает интенсивно засасывать окружающий воздух всей боковой поверхностью колонны. В данной работе не решается проблема раскручивания столба вокруг своей оси. Примем этот факт как данный. В результате установившегося вращения на границе колонны должна образоваться тонкая быстро вращающаяся граница в виде трубчатой водяной плёнки. В момент ее образования из отдельных капель шнур должен резко сжаться из-за возникновения СПН. Граничные условия на такой границе должны удовлетворять условию: $\Delta \cdot \rho_s V_\vartheta^2 \sim D\sigma$ или $\Delta \sim \dfrac{D\sigma}{\rho V_\vartheta^2}$, где $\rho_s$ - плотность массы вращающегося слоя колонны, $\Delta$ - толщина этого слоя. Стенка такой трубы состоит из водяной плёнки. При установившейся скорости вращения колонны, измеряемая линейная скорость вращения, её пропитанной водой стенки $V_\vartheta \sim 10^4$ см/с, т.е., порядка скорости вдоль оси. Отсюда получаем $\Delta \sim 6$ см. Если же этот слой вспенен пузырьками воздуха, засасываемого в слой, то толщина этого слоя будет ещё больше. Но для оценки количества воды, засасываемого через этот слой МО, мы воспользуемся именно этой цифрой. При этих предположениях объём засасываемой воды МО за секунду равен ~ 10⁸ см³. За треть час эта величина составит ~ 10¹¹ см³. Если площадь МО равна ~ 25 км² (что относится к реальным размерам для ААЯ ВС) и вся захваченная за треть часа МО вода вновь возвращается на землю в виде осадков на площадь равную площади самого облака, то средний уровень этих осадков составит величину ~ 0,4 см. На самом деле уровень осадков будет выше, поскольку и другая часть площади колонны поставляет влагу в МО, но эта часть не может на порядки превышать данную оценку. Такая же оценка массы воды получается при оценке через поток капель. Она оказывается ~2 10⁸ см³/сек. Оценим теперь энергетику данного явления по работе совершённой в этом процессе подъёма воды на высоту порядка 1 км в течение трети часа. Принимая во внимание полученные выше оценки, получаем, во-первых, значение мощности этого процесса $W \sim 10^9$ Ватт. За треть часа работы такая колонна затрачивает ~ 4 10¹² Дж. Эта цифра порядка той, которая характеризует классический торнадо. Совершенно очевидно, что без наличия капиллярной составляющей в воде при наблюдаемых значениях потенциала такого процесса образования ААЯ ВС быть не может. Действительно, при исходных значениях потенциала не происходит такого накопления плотности заряда на плоской поверхности воды, которая привела бы к отрыву от нее капель или взорвала бы её за счет кулоновского взрыва.

**4. Заключение.** Сравним теперь наблюдаемые явления, сопровождающие возникновение этого ААЯ, с ожидаемыми проявлениями по предложенной модели ВС, исключая пока все, что связано с природой и развитием вращения. Начнем с совпадений. 1. Начальная стадия развития ВС, по предложенной модели, не требует вращения его колонны, что и наблюдает-ся. 2. По модели в зоне отрыва капель от капилляров должно наблюдаться свечение. Оно обусловлено возбуждением молекул воды и воздуха, а также процессами их ионизации/рекомбинации. Таким образом, зона отрыва капель по модели соответствует светлому пятну на воде при зарождении реального ВС. 3. При установившемся вращении колонны, в момент образования замкнутой пленки жидкости на ее периферии, шнур должен резко сократиться в диаметре, из-за возникших СПН, которые с этого момента уравновешивают центробежную силу в колонне. Это соответствует реальным наблюдениям. 4. На стадии ВС с вращением колонны она имеет трубчатое строение, как и должно быть по предложенной модели. 5. Суммарная энергетика и количество транспортируемой в МО воды ВС по предложенной модели соответствует наблюдаемым величинам. 6. Согласно представленной гипотезе, процесс перехода ВС в классический торнадо идентичен наблюдаемому процессу такого развития. В дальнейшем автору предстоит описать процесс раскручивания колонны ВС.